\documentclass[twocolumn,showpacs,preprintnumbers,amsmath,amssymb]{revtex4}
\usepackage{graphicx}
\usepackage{dcolumn}
\usepackage{bm}
\begin{document}

\title{Diffusion of Nonequilibrium Quasiparticles in a Cuprate Superconductor}
\author{N. Gedik}
\email{gedik@socrates.berkeley.edu}
\author{J. Orenstein}
\affiliation{Physics Department, University of California,
Berkeley and \\ Materials Science Division, Lawrence Berkeley
National Laboratory, Berkeley, CA 94720}
\author{Ruixing Liang}
\author{D.A. Bonn}
\author{W.N. Hardy}

\affiliation{Department of Physics and Astronomy, University of British Columbia, Vancouver, British Columbia, Canada V6T 1Z1 }%

\begin{abstract}
We report a transport study of nonequilibrium quasiparticles in a
high-$T_c$ cuprate superconductor using the transient grating
technique.  Low-intensity laser excitation (at photon energy 1.5
eV) was used to introduce a spatially periodic density of
quasiparticles into a high-quality untwinned single crystal of
YBa$_2$Cu$_3$O$_{6.5}$. Probing the evolution of the initial
density through space and time yielded the quasiparticle diffusion
coefficient, and both inelastic and elastic scattering rates. The
technique reported here is potentially applicable to precision
measurement of quasiparticle dynamics, not only in cuprate
superconductors, but in other electronic systems as well.

\end{abstract}
\pacs{74.25.Gz, 78.47.+p}
\maketitle


Quasiparticles are the elementary excitations of a superconductor,
created when a Cooper pair of electrons breaks apart.  The dynamic
properties of quasiparticles, that is their rates of diffusion,
scattering, trapping, and recombination, are critical for
applications of conventional superconductors in X-ray detectors
\cite{xray} and in the manipulation of superconductor-based qubits
\cite{qubit}.  In more exotic superconductors, such as the
high-$T_c$ cuprates, a better understanding of quasiparticle
dynamics may help to uncover the mechanism for Cooper pairing.  A
special property of the cuprate superconductors is the d-wave
symmetry of the gap function, which leads to an unusual
quasiparticle spectrum. The minimum energy for the creation of a
quasiparticle depends on the direction of its momentum
\cite{shen}. It is zero for momenta in the 'nodal' direction,
oriented at 45$^\circ$ relative to the Cu-O bond. The most
energetically expensive quasiparticles are the 'antinodal' ones,
whose momenta are nearly parallel to the bond. The antinodal
quasiparticles are the mystery particles of cuprate
superconductivity.  Because they feel the pairing interaction most
strongly, their properties may hold the key to high-$T_c$
superconductivity.  Unfortunately, their tendency to form strong
pairs makes them difficult to study.  In thermal equilibrium the
population of quasiparticles is overwhelmingly dominated by the
low energy nodal ones.  As a result, transport measurements
performed in equilibrium, such as microwave \cite{hosseini} and
thermal \cite{ong} conductivity, are insensitive to antinodal
quasiparticles.

In this work a transient grating technique was developed and used
to probe the transport of nonequilibrium quasiparticles in the
high-$T_c$ superconductor YBa$_2$Cu$_3$O$_{6.5}$.  We find that
the diffusion coefficient, $D$, is much smaller than the value
obtained in measurements on equilibrium quasiparticles in the same
material. The disparity in $D$ suggests that the current
experiment probes a population of quasiparticles that are not near
the nodes, and are perhaps close to the antinodal regions of
momentum space.

The nonequilibrium quasiparticles were introduced using short
optical pulses. To probe their propagation, we generated a
spatially periodic population by interfering two pulses at the
sample surface. The spatial period, $\lambda_g$, equals
$\lambda/2$sin$\theta$, where $\lambda$ is the wavelength of the
pulse and $2\theta$ is the angle between the two pump beams. The
nonequilibrium quasiparticles cause a change in the index of
refraction at the laser frequency \cite{chwalek} which is a linear
function of their density \cite{segre}. As a result, the
sinusoidal variation in quasiparticle density creates an index
grating which can be detected by the diffraction of a probe pulse.

After creation, the distribution of quasiparticles evolves due to
the combined effects of recombination and diffusion. In the
process of recombination a pair of quasiparticles jumps back into
the Cooper pair condensate with the simultaneous transfer of their
creation energy to some other form (e.g. phonons).  The amplitude
of the grating may also decay as quasiparticle diffusion drives
the system towards a spatially homogeneous quasiparticle
concentration.  The goal of the experiment is to disentangle these
effects and measure both the rates of recombination and diffusion.

The transient grating technique has been used successfully in a
wide variety of applications, including exciton diffusion,
dynamics of biomolecules, propagation of ultrasound, and thermal
diffusion \cite{tgbook}. Observing the propagation of
superconducting quasiparticles requires the ability to detect the
transient grating at extremely dilute concentrations.  At the low
excitation densities needed to detect their propagation, the
quasiparticles produce no more than $10^{-5}$ fractional change in
the index of refraction.  The diffracted intensity from such a
grating would therefore be of order $10^{-10}$ of the incident
probe intensity and consequently very difficult to detect.

Detection of the grating required measurement of the amplitude of
the diffracted wave, which is a part in $10^5$ of the probe,
rather than the intensity. This is accomplished through the use of
heterodyne detection \cite{heterodyne1,heterodyne2}.  The key
element in the success of the heterodyne technique is the
diffractive optic (DO) beamsplitter, which creates pairs of pump
and probe beams \cite{DO1,DO2}. The DO element in our experimental
setup (Fig. 1) is an array of 10 separate 2.5 mm square phase
masks, each with a different grating period, on a fused silica
substrate. A beam from a Ti:Sapphire laser ($\lambda$=800 nm and
pulse repetition rate 80 MHz) is split into primary pump and probe
beams, which are focused onto one of the phase masks at a small
angle with respect to each other.  The phase mask splits each of
the primary beams by diffraction into the m=$\pm$1 orders. (For
clarity, only the probe beam paths are shown in Fig. 1).

The interference of the two pump beams creates a spatially varying
index of refraction in the sample which reaches a depth 1000 $\AA$
below the surface and has half the period of the phase mask. We
detect the index variation using an implementation of the
heterodyne technique that enables absolute calibration of the
phase \cite{SOM}. Each of the two probe beams is specularly
reflected from the surface of the sample, and returns, via the DO,
to a Si detector. The two possible round-trip beam paths are shown
as solid and dashed lines in Fig. 1.  With the use of the DO
beamsplitter, detection of the diffracted probe is automatically
aligned.  The diffracted component of P1 is precisely colinear
with reflected P2, and vice versa.  The experiment is performed by
alternately blocking one of the two reflected beams.  If reflected
P1 is blocked, then reflected P2 and diffracted P1 are mixed in
the detector.  This measurement by itself is insufficient to
extract the wave amplitudes because the relative phase of P1 and
P2 is undetermined.  However, simply blocking reflected P2 instead
of P1 produces a mixed signal with the conjugate phase.  Comparing
the detector output for the two conjugate beam paths fixes the
absolute phase and therefore the wave amplitudes as well.

\begin{figure}[tb]
\includegraphics[width=3.25in]{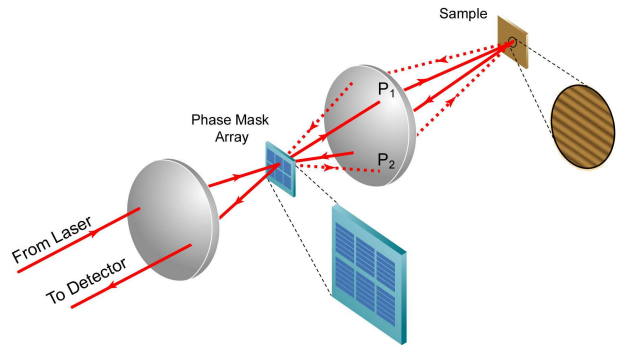}
\caption{Illustration of the beam path for heterodyne transient
grating detection. Pump and probe beams from the laser are split
at the diffractive optic (for clarity only the probe beams are
shown). A spherical mirror and plane folding mirror (represented
schematically by a lens in the sketch) focus the beams to a single
100 $\mu$m spot on the sample.  After specular reflection and
diffraction at the sample surface, the two probe beams are
recombined by the diffractive optic and directed to a Si
photodiode detector. The wavevector of the quasiparticle density
variation is changed, without optical realignment, by translating
the diffractive optic so that a different phase mask in the array
is inserted in the beam.}

\end{figure}

By adjusting the phase delay of P1 relative to P2, the output of
the Si photodiode measures either the change of the specular
reflection coefficient due to the grating, \textit{R}, or the
amplitude of the diffraction efficiency, \textit{TG}.  \textit{R}
is proportional to the spatial average of the quasiparticle
concentration, whereas \textit{TG} is proportional to the
component of the quasiparticle concentration at the fundamental
period of the grating, $\lambda_g$ \cite{fayer}. In Fig. 2 we plot
both \textit{R} and \textit{TG} for $\lambda_g=$2 $\mu$m as a
function of time delay, for several intensities of the excitation
pulses, and consequently for a range of initial quasiparticle
concentrations. The temperature of the sample was 5 K and the
grating wavevector was oriented along the crystallographic
\textbf{b} axis (parallel to the Cu-O chains).

\begin{figure}[tb]

\includegraphics[width=3.25in]{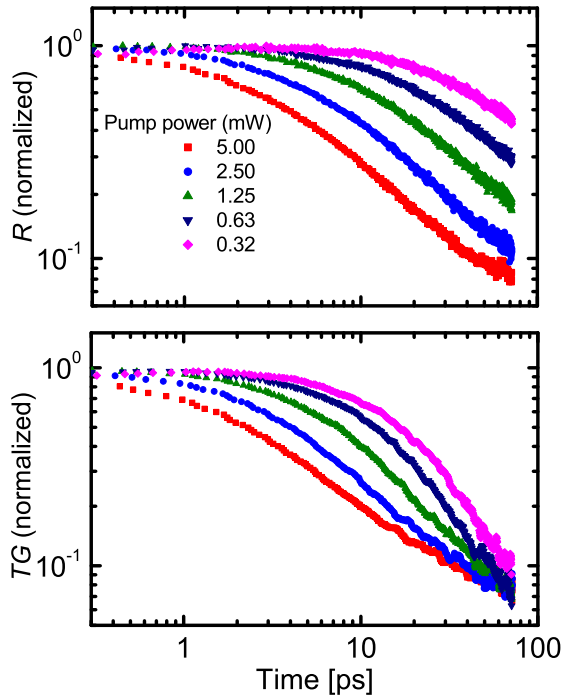}
\caption{(top panel) Change in the specular reflection
coefficient, \textit{R}, as a function of time delay following
creation of the grating, for several values of the pump intensity.
\textit{R} is normalized to unity at time delay zero to illustrate
the systematic slowing down of the recombination rate as the
excitation density is decreased. (bottom panel) Normalized
amplitude of the diffracted probe beam as a function of time delay
for the same values of pump intensity used to measure \textit{R}.}

\end{figure}
The curves are normalized to their value at time delay zero. The
\textit{R} curves, which decay due to recombination, are
nonexponential, and their characteristic rate of decay increases
with increasing concentration \cite{segre}.  The increase of
recombination rate with density is consistent with the idea that
each quasiparticle must encounter another to scatter into the
Cooper pair condensate.  The corresponding \textit{TG} curves
depend strongly on the pump intensity as well.  However the time
dependence at each pump intensity is different from \textit{R}
because \textit{TG} reflects the combined effects of recombination
and propagation. In systems where the recombination rate is
independent of density, it is relatively straightforward to
separate the effects of particle decay and diffusion.  If the
average concentration decays exponentially with rate $\gamma$, the
amplitude of the grating decays with rate $\gamma+Dq^2$, where
\textit{D} is the diffusion coefficient and \textit{q} is the
wavevector of the grating, $2\pi/\lambda_g$.  The ratio
\textit{TG/R}, which decays simply as exp(-$Dq^2t$), isolates the
effects of diffusion on the evolution of the particle density.

With this example in mind, we are led to consider \textit{TG/R}
for our data (see Fig. 3).  The left, and right, panels show
\textit{TG/R} for grating periods of 2, and 5 $\mu$m,
respectively, for several values of the pump laser intensity.
Unlike the example of density-independent recombination, the decay
of \textit{TG/R} is nonexponential and highly dependent on the
excitation density. At high intensity \textit{TG/R }recovers its
value at zero time delay, after an initial rapid decrease.
\textit{TG/R }is nearly independent of grating period at high
intensity.  As intensity decreases, the minimum of \textit{TG/R}
moves systematically towards longer times and the decay approaches
a simple exponential.  The exponential rate at low pump laser
intensity clearly depends strongly on $\lambda_g$, as is expected
if the grating decays due to diffusion.

\begin{figure}[tb]
\includegraphics[width=3.25in]{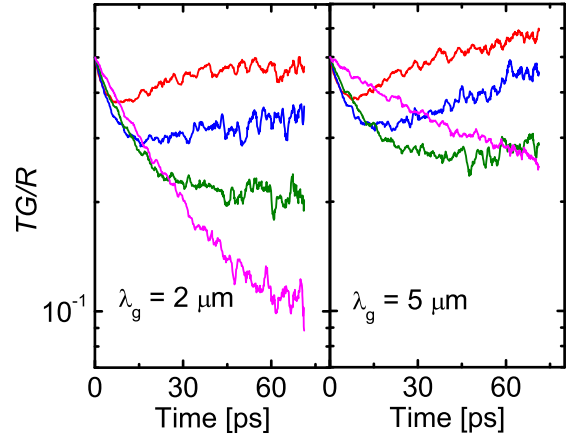}
\caption{The ratio \textit{TG/R} for the same pump intensities as
in Fig. 2 (with P=0.63 mW omitted for clarity) for grating period
2$\mu$ (left panel) and 5$\mu$ (right panel). At high excitation
density the curves are nearly independent of the period.  At low
excitation density the grating enters the propagation-dominated
regime where the decay of \textit{TG/R} depends strongly on the
period of the grating.}

\end{figure}

The dependence of \textit{TG/R} on laser intensity and grating
period results from an interplay of density-dependent
recombination, diffusion, and energy transfer. To unravel these
effects we have modeled the quasiparticle dynamics by adding a
quadratic recombination term to the diffusion equation: $\partial
n(x,t)/\partial t = D\partial^2 n(x,t)/\partial x^2-\beta
n^2(x,t)$ \cite{SOM}. Here, \textit{n} is the quasiparticle
density and $\beta$ is the recombination coefficient, which is a
measure of the inelastic scattering rate \cite{scalapino}. The
dynamics predicted by this equation depend on the relative
magnitude of $\beta n$ and $Dq^2$. In the high density regime,
$\beta n \gg Dq^2$, recombination dominates over diffusion.
Because the rate is more rapid where \textit{n} is larger,
recombination distorts the sinusoidal grating by flattening the
crests.  The distortion of the grating profile reduces the
component of \textit{n} at the fundamental grating period faster
than its spatial average.  This accounts for the initial decrease
of \textit{TG/R}. However, if this were the only process
\textit{TG/R} would never bounce back to its initial value, as it
is seen to do in Fig. 3.

\textit{TG/R} recovers in the high-intensity regime because the
energy released as the quasiparticles recombine appears in another
form, one that also creates a change in the index of refraction.
In conventional superconductors the quasiparticle energy is
converted to lattice vibrational energy, which (in the presence of
electron-phonon coupling) will change the index. While phonons may
play this role in the cuprate superconductors as well, excitations
involving the flipping of electron spins may be generated instead
when quasiparticles recombine \cite{quinlan}. In either case, the
recovery of \textit{TG/R} follows if this secondary form of energy
does not diffuse on the 100 ps time scale.  In this case, the
original sinusoidal distribution of energy is written into a
stationary form of energy before the quasiparticles have had a
chance to diffuse. When almost all the energy has been handed off
from the quasiparticles to non-propagating modes, the grating
recovers its sinusoidal profile and \textit{TG/R} returns to its
value at time zero. The energy remains frozen until the nanosecond
time scale, when thermal diffusion becomes significant
\cite{fayer2}.

The effects of diffusion dominate in the low density regime,
$\beta n \ll Dq^2$, where quasiparticle motion washes out the
grating before energy transfer can take place.  According to the
equation governing $n(x,t)$, \textit{TG/R} becomes \textit{q}
dependent and intensity independent in this limit.  To find the
rate of diffusion, we measured \textit{TG/R} at low power (P=0.32
mW) for several grating periods between 2 and 5 $\mu$m. The
initial decay rate of \textit{TG/R} at low power is plotted as a
function of $q^2$ in Fig. 4. The decay rate depends systematically
on the grating period, demonstrating that the dynamics are in the
propagation-dominated regime. The rates for the grating oriented
along both the \textbf{a} and \textbf{b} crystallographic
directions are plotted. For both directions the rate is a linear
function of $q^2$, demonstrating that the quasiparticle
propagation is diffusive. From the slope of a linear fit we
determine that $D_a$=20 cm$^2$/s and $D_b$=24 cm$^2$/s. The
intercept as \textit{q} tends to zero is the decay rate due to
recombination.

\begin{figure}[tb]
\includegraphics[width=3.25in]{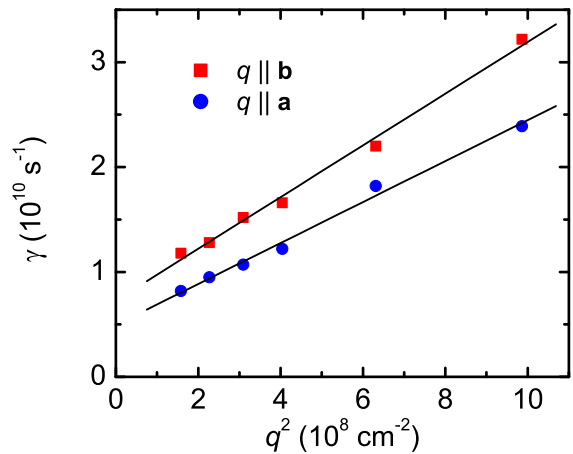}
\caption{Initial decay rate of \textit{TG/R}, for the same low
intensity as in Fig. 3, plotted as a function of the square of the
grating wavevector.  Results are shown for two perpendicular
orientations of the grating.  The linear dependence of the rate on
$q^2$ demonstrates that the propagation is diffusive. The slopes
of linear fits yield diffusion coefficients $D_a$ = 20 cm$^2$/s
and $D_b$ = 24 cm$^2$/s.}

\end{figure}

It is possible to infer the mean free time $\tau$ and mean square
velocity $<v^2>$ of nonequilibrium quasiparticles from the values
of $D$ quoted above. A lower bound on $\tau$ is obtained by
inserting the maximum quasiparticle velocity, the Fermi velocity
$v_F$, into the kinetic formula, $D=<v^2>\tau/2$. The literature
value \cite{valla} $v_F=2\times10^7$ cm/s yields a lower bound of
100 fs. An upper bound on $\tau$ of essentially the same value can
also be inferred from the experiment. If $\tau$ were significantly
longer than 100 fs then quasiparticle propagation would be
ballistic on the subpicosecond time scale. However, the time and
wavevector dependence of $TG$ proves that quasiparticle motion is
diffusive at the earliest times we can resolve, which is $\sim$
300 fs after creation of the grating. Thus the allowed values of
$\tau$ and $v$ are narrowly bracketed near 100 fs and $v_F$,
respectively. Furthermore, the measurements indicate that $\tau$
is determined by elastic, rather than inelastic, processes. The
nonequilibrium particles survive for $\approx$ 100 ps, and
therefore scatter $\approx$ 1000 times before decaying. This would
be impossible if each scattering event resulted in a significant
reduction of the quasiparticle's energy.

Equilibrium measurements find (in the same crystal and at the same
temperature) that $\tau=$ 20 ps \cite{turner}, which is 200 times
longer than $\tau$ of the nonequilibrium quasiparticles. The
contrast suggests that the nonequilibrium quasiparticles are
different from those present in thermal equilibrium. The
equilibrium particles occupy states which have a small energy
($k_BT$) relative to the chemical potential, and therefore lie
very near the gap nodes. If the nonequilibrium quasiparticles are
different, they must occupy higher energy states, perhaps ones
closer to the antinodal regions of momentum space. It is possible
that the constraints of momentum and energy conservation prevent
relaxation of antinodal quasiparticles to the nodal regions. If
the particles are antinodal, it is relevant to compare the width
of the antinodal quasiparticle peak as measured by photoemission
with $\hbar/\tau$ as measured by nonequilibrium transport. The
peak width of 14 meV \cite{fedorov} corresponds to $\tau$=50 fs,
which is close to the transport value, particularly considering
that photoemission is performed on Bi$_2$Sr$_2$CaCu$_2$O$_8$
rather than YBa$_2$Cu$_3$O$_{6.5}$. The similarity of lifetimes
suggests that the peak width in photoemission may be controlled by
elastic scattering as well.

The transient grating method reported here promises to be broadly
applicable to superconductors, as well as other materials in which
there is a gap in the quasiparticle spectrum.  The technique works
readily in transmission or reflection geometry and therefore can
be applied to bulk materials or thin films.  The propagation of
quasiparticles can be tracked in any system where nonequilibrium
excitations generate a change in the index of refraction at the
laser wavelength.  In conventional superconductors, quasiparticle
diffusion can be measured without fabricating trapping layers and
junction detectors.  In more exotic systems with multiple or
anistropic gaps, such as reported in this paper, the transient
grating technique can track the propagation of quasiparticles that
conventional transport methods cannot detect.

This work was supported by DOE-DE-AC03-76SF00098, CIAR, and NSERC.

\end{document}